\documentclass[journal,twoside]{IEEEtran}
\usepackage{cite,color}
\usepackage[normalem]{ulem}
\usepackage{algorithm}
\usepackage{algpseudocode,subfigure,mathtools}

\ifCLASSINFOpdf
  \usepackage[pdftex,final]{graphicx}
\else
  \usepackage[dvips,final]{graphicx}
  \DeclareGraphicsExtensions{.eps}
\fi
\usepackage[outdir=./]{epstopdf}
\usepackage{amsfonts}
\usepackage{amsmath}
\usepackage{amsthm}
\usepackage[cmintegrals]{newtxmath}
\usepackage{breqn}

\usepackage{array}
\newtheorem{lemma}{\underline{Lemma}}

\newtheorem{remark}{\underline{Remark}}

\DeclareMathOperator*{\argmax}{arg\,max}

\newcommand{\norm}[1]{\lVert {#1} \rVert}

\begin{document}
\title{A New View of Multi-User Hybrid Massive MIMO: Non-Orthogonal Angle Division Multiple Access}
\author{Hai~Lin,~\IEEEmembership{Senior Member,~IEEE}
, Feifei~Gao,~\IEEEmembership{Senior Member,~IEEE}
, Shi~Jin,~\IEEEmembership{Member,~IEEE}
, Geoffrey~Ye~Li,~\IEEEmembership{Fellow,~IEEE}
\thanks{H. Lin is with the Department of Electrical and Information Systems, Graduate School of Engineering, Osaka Prefecture University, Sakai, Osaka, Japan  (e-mail: hai.lin@ieee.org).} 
\thanks{F. Gao is with the Department of Automation, Tsinghua University, State Key Lab of Intelligent Technologies and Systems, Tsinghua National Laboratory for Information Science and Technology (TNList) Beijing, P. R. China (e-mail: feifeigao@ieee.org).}
\thanks{S. Jin is with the National Communications Research Laboratory, Southeast University, Nanjing 210096, P. R. China (email: jinshi@seu.edu.cn).}
\thanks{Geoffrey Ye Li is with the School of Electrical and Computer Engineering, Georgia Institute of Technology, Atlanta, GA, USA (email: liye@ece.gatech.edu)}
}

\maketitle

\begin{abstract}
This paper presents a new view of multi-user (MU) hybrid  massive multiple-input and multiple-output (MIMO) systems from array signal processing perspective.
We first show that the instantaneous channel vectors corresponding to different users are asymptotically orthogonal if the angles of arrival (AOAs) of users are different.
We then decompose the channel matrix into an angle domain basis matrix and a gain matrix. The former can be formulated by steering vectors and the latter has the same size as the number of RF chains, which  perfectly matches the structure of hybrid precoding. A novel hybrid channel estimation is proposed by separately estimating the angle information and the gain matrix, which could significantly save the training overhead and substantially improve the channel estimation accuracy compared to the conventional beamspace approach. Moreover, with the aid of the angle domain matrix, the MU massive MIMO system can be viewed as a type of non-orthogonal angle division multiple access (ADMA) to simultaneously serve multiple users at the same frequency band. Finally, the performance of the proposed scheme is validated by computer simulation results.
\end{abstract}

\begin{IEEEkeywords}
Massive MIMO,  array signal processing, angle of arrival (AOA), channel estimation, hybrid precoding, angle division multiple access (ADMA).
\end{IEEEkeywords}

\IEEEpeerreviewmaketitle

\section{Introduction}
Massive multiple-input and multiple-output (MIMO) has been considered as a key technology for the next-generation cellular system\cite{ng,chen1,chen2}, where multiple users (MU) are simultaneously served at the same frequency band by the base station (BS) equipped with a large number of antennas.
The benefits brought by massive BS antennas include high energy efficiency, high spectrum efficiency, high spatial resolution, broad  coverage, and so on\cite{scaling,overview}.
For MU massive MIMO systems,  downlink (DL) transmission  relies on the precocding to reduce inter-user interference (IUI).  In time division duplex (TDD) systems where channel reciprocity holds, the channel state information (CSI) required by DL precoding can be obtained via uplink (UL) pilots to avoid the significant  training overhead \cite{tdd}. Clearly, the UL channel estimation and the DL precoding are the foundation of the TDD massive MIMO systems.

When the number of the antennas approaches infinity, channel vectors corresponding to different users are spatially orthogonal and the optimal DL precoding is simply the matched filtering. However, if there are a finite number of antennas in practical systems, such an orthogonality does not hold and the massive MIMO system naturally turns into a non-orthogonal multiple access (NOMA) system. Therefore, more sophisticated precoding schemes are necessary.  The concept of NOMA has been originally proposed to enhance the spectrum efficiency by allowing multiple users allocated with different power levels to share the same resource block \cite{NOMA01,NOMA1,NOMA2,NOMA3}. Another dominant NOMA category is code-domain multiplexing, including multiple access low-density spreading CDMA \cite{NOMA4,NOMA5}, sparse-code multiple access \cite{NOMA6}, multi-users shared access \cite{NOMA7}, and so on. Several other multiple access schemes have also been put forward, such as pattern-division multiple access, bit-division multiplexing\cite{NOMA8}, and interleave-division multiple access\cite{NOMA9}. An earlier study of spatially non-orthogonal multiple access scheme has been presented in \cite{NOMA10} for conventional MIMO where multiple users are served by non-orthogonal beams. Like other NOMA systems, understanding the rationale behind the channel non-orthogonality in massive MIMO   will definitely benefit the system design.

Another critical issue  of massive MIMO is the practical cost associated with a large number of RF chains.
As a cost-effective solution, massive antennas with limited RF chains have attracted  substantial  attention, where the precoding is performed in a hybrid manner by combining phase shifters based analog precoding and a much smaller size  digital precoding. Since its first appearance under the name of antenna soft selection \cite{molish}, hybrid precoding has been studied extensively, as can be seen from  \cite{yu} and the references therein. However, it is still unclear how to well form analog precoding, especially for low-cost phase shifters with \emph{finite resolution}.

Meanwhile, most existing hybrid precoding schemes depend on the perfect knowledge of the channel. However, when there are only limited RF chains available, channel estimation becomes challenging.
An adaptive algorithm to estimate the hybrid millimeter wave (mmWave) channel parameters has been developed in \cite{HB1}, where the poor scattering nature of the channel is exploited and adaptive compressed sensing (CS) has been employed.
The accuracy of the CS method is limited by the finite grid and its computational complexity is also high for the practical deployment.
A beam training procedure has been provided in \cite{HB2}, which aims to search only several candidate beam pairs for fast channel tracking. Although this category of schemes works well for the point-to-point scenarios, the pilot overhead is very high for the multi-user scenarios.
A beam cycling method has been developed in \cite{cycling}, where channel estimation comparable to  the full digital system is achieved by sweeping the beam directions over all spatial region.
A prior knowledge aided hybrid channel tracking scheme has been developed in \cite{HB4} for Tera-Hertz beamspace massive MIMO systems, which excavates a temporal variation law of the physical direction for predicting the support of the beamspace channel. However, the prior knowledge of the user is not always known in practice and the method cannot be applied for the more general case.
Recently, an array signal processing aided channel estimation scheme has been proposed in \cite{sbem,sbem1,sbem2},
where the angle information of the user is exploited to simplify the channel estimation.
Nevertheless, the scheme is only applicable for full digital systems and cannot be directly extended to hybrid systems.

In this paper, we develop a novel hybrid massive MIMO transmitter from the angle domain perspective, which could significantly save the training overhead and substantially improve the channel estimation accuracy compared to the conventional beamspace approach. We first analyze instantaneous channels in a massive MIMO system, where the BS is equipped with an $M$-antenna uniform linear array (ULA) and each user has single antenna.
It is shown that the channel vectors corresponding to different users are asymptotically orthogonal as $M$ goes large, when the angles of arrival (AOAs) of users are different.
Using the discrete Fourier transform (DFT), the \emph{cosine} of the AOA can be estimated with a resolution proportional to $1/M$. The resolution can be further
enhanced by using \emph{zero padding} technique with fast Fourier transform (FFT).
We then decompose the channel matrix into an angle domain basis matrix and a corresponding gain matrix.
The former can be formulated either by the orthogonal or the non-orthogonal steering vectors and the latter has the same size as RF chains.
Accordingly, the precoding scheme consists of either orthogonal or non-orthogonal beamforming towards users and an angle domain precoding dealing with the IUI.
By mapping the above beamforming and precoding matrices to the analog and the digital precoding, respectively,
the proposed scheme perfectly matches the hybrid precoding with finite resolution phase shifters.
From the AOA-based analysis, the MU massive MIMO  can be viewed as an angle division multiple access (ADMA) system.

The rest of the paper is organized as follows: Section II investigates the channel orthogonality from the viewpoint of AOA, and then training-based AOA estimation is discussed in Section III. Next, angle-domain decomposition aided hybrid precoding is proposed in Section IV, followed by a novel hybrid channel estimation scheme in Section V. Simulation results are presented in Section VI, and Section VII concludes the paper.

\section{Channel Characteristics}

Consider an MU massive MIMO system, where the BS is equipped with  a ULA of $M$ elements to serve $K$ single-antenna users.
From the well-established narrowband transmission model\cite{aoamodel}, the UL channel vector between the $k$th user and the BS can be expressed as
\begin{eqnarray}\label{hk}
{\mathbf h}_k=\sum_{p=1}^{P}{\mathbf a} (\theta_{k,p}) \frac{\alpha_{k,p}}{\sqrt{P}},
\end{eqnarray}
where $P$ is the number of i.i.d paths,
$\alpha_{k,p}\sim \mathcal{CN}(0,\sigma^2_k)$ is the complex gain of the $p$th path of the $k$th user,
and ${\mathbf a} (\theta_{k,p})$ is the steering vector. The steering vector can be expressed as
\begin{eqnarray}
{\mathbf a} (\theta_{k,p})
=
\left[
\begin{array}{c}
 1\\
 e^{-j2\pi \frac{D}{\lambda} \cos(\theta_{k,p})} \\
 \vdots \\
e^{-j2\pi (M-1) \frac{D}{\lambda} \cos(\theta_{k,p})}
\end{array}
\right],
\end{eqnarray}
where $D \le \lambda/2$ is the BS antenna spacing,  $\lambda$ is the wavelength of the carrier frequency, and $\theta_{k,p} \in [0, \pi]$ is a random AOA.
The matrix form of (\ref{hk}) can be written as
\begin{eqnarray}
\mathbf h_k  =  \mathbf A_k \boldsymbol \alpha_k, \label{eq:hkmatrix}
\end{eqnarray}
where $\mathbf A_k=[{\mathbf a} (\theta_{k,1}), \ldots, {\mathbf a} (\theta_{k,P})]$ and $\boldsymbol \alpha_k=[\alpha_{k,1}/\sqrt{P},\ldots,\alpha_{k,P}/\sqrt{P}]^T$.
Then, based on channel model in (\ref{eq:hkmatrix}), we can prove the following lemma,
which will be very useful to the design of an MU massive MIMO system.

\begin{lemma}
For any two channel vectors, ${\mathbf h}_k$ and ${\mathbf h}_{n}$ $(k \ne n)$, if they have no common path, namely $\theta_{k,i}\ne \theta_{n,j}$ for any $i$ and $j$, then they are asymptotically orthogonal when $M$ is sufficiently large. That is,
$\displaystyle{\lim_{M\rightarrow \infty}} \gamma(k,n)=0$, where $\gamma(k,n)=\left(\frac{\mathbf h_k}{\norm{\mathbf h_k}}\right)^{\mathcal H}\left(\frac{\mathbf h_n}{\norm{\mathbf h_n}}\right)$.
\end{lemma}
\begin{IEEEproof}
See Appendix A.
\end{IEEEproof}

The result in the Lemma 1 does not rely on any statistical properties of the channel tap gain vectors $\boldsymbol \alpha_k$ and $\boldsymbol \alpha_n$, and thus can directly apply to instantaneous channels.
Consequently, when the BS array is large enough and the channels do not share the same paths, they are always orthogonal regardless of whether  their fading coefficients are correlated or not. On the other hand, when there are common paths between two instantaneous channels, the orthogonality depends on their instantaneous fading coefficients\cite{tdd},\cite{Erik}.

\begin{figure*}[!t]
\centering
\includegraphics[width=11cm]{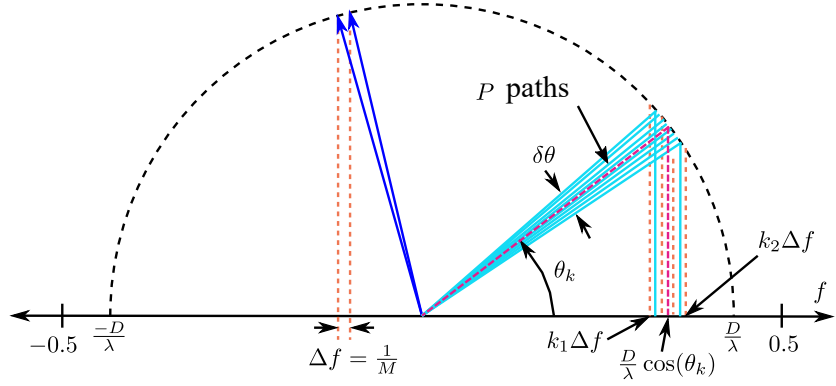}
\caption{The relation between $\theta$ and $f$, $D\le \lambda/2$.}
\label{aoa_narrow_as}
\end{figure*}

\section{Training based AOA Estimation}

From Section II, the AOA plays a fundamental role in the analysis of massive MIMO system and thus needs to be obtained.
AOA estimation is a classical problem in the area of array signal processing, and there are many well-known approaches.
For example, the parametric algorithms MUSIC \cite{MUSIC} and ESPRIT \cite{ESPRIT} have already demonstrated  their superior resolution capability compared to the non-parametric counterparts.
However, most existing AOA estimation methods are based on the statistics of the received signal, say the signal covariance, because traditional AOA estimation is \emph{passive} and blind, and the targets do not
want to be detected and do not have any interplay with the array.

On the other hand, in the area of wireless communications,
the users want to be connected by the BS array and therefore could help the BS estimate their AOAs by sending pilot signals.  For example, $K$ users can send $K$ orthogonal training sequences of length $K$ to the BS to estimate the instantaneous UL channel vector, $\mathbf h_k$, for each user \cite{channelesti1}. The next question is how AOA is related to or can be obtained from $\mathbf h_k$.

Denote $f_{k,p}=\frac{D}{\lambda}\cos(\theta_{k,p})$. Then $\mathbf A_k$ in (\ref{eq:hkmatrix}) can be expressed as
\begin{eqnarray}\label{Ak}
\mathbf A_k=
\left[
\begin{array}{ccc}
 1 & \ldots & 1 \\
 e^{-j2\pi f_{k,1}} & \ldots & e^{-j2\pi f_{k,P}} \\
 \vdots & \vdots & \vdots \\
e^{-j2\pi (M-1)f_{k,1}} & \ldots & e^{-j2\pi (M-1)f_{k,P}}
\end{array}
\right].
\end{eqnarray}
From (\ref{Ak}), the AOA estimation is equivalent to the well-known \emph{multiple-frequency estimation} problem, where the {\it frequencies} to be estimated are $f_{k,p}$. It is obvious that $|f_{k,p}|\le 0.5$ since $D/\lambda \le 0.5$. In the following discussion, we let $D=\lambda/2$, for the sake of convenience.

A simple way is to apply the DFT on $\mathbf h_k$, which can be implemented by the FFT if $M$ is power of $2$.  The resolution of the estimation is $\Delta f= 1/M$. Therefore,
it may not be possible to recover all the angle and gain parameters if the number of antennas is finite. In other words, two ``frequencies" with a difference less than $1/M$ can not be distinguished if there are $M$ antennas at the BS. Fortunately, we do not need all these $P$ AOAs of the same user for the hybrid precoding design, as we can see later.
For a massive MIMO system, i.e., $M\rightarrow \infty$, it is possible to recover all $f_{k,p}$ as well as $\alpha_{k,p}$ since there are much more observations than unknown variables.

As in Fig.\ref{aoa_narrow_as}, we have the following assumption:
\emph{
The incoming AOAs of any user are restricted in a small region, i.e.,
$\theta_{k,p}\in [-\frac{\delta\theta}{2}+\theta_k, \frac{\delta\theta}{2}+\theta_k]$ for $1\le p \le P$, where
the angular spread $\delta\theta\ll \pi$.} This happens when the BS is installed on very tall building and there are a limited number of surrounding scatterings\cite{adhikary} or when mmWave scenario is exploited\cite{cycling}.

With the above assumption, $\theta_{k,p}$'s are very close to each other and ${\mathbf a} (\theta_{k,p})$'s are highly correlated. Hence, $\mathbf h_k$ in (\ref{eq:hkmatrix}) can be re-expanded more compactly by a much smaller number of steering vectors. Denote

\begin{eqnarray}\label{FM}
\mathbf {F}=\frac{1}{\sqrt{M}}
\left[
\begin{array}{cccc}
1 & 1 & \ldots & 1 \\
1 & e^{-j2\pi \Delta f} & \ldots & e^{-j2\pi (M-1)\Delta f} \\
\vdots & \vdots & \vdots & \vdots \\
1 & e^{-j2\pi (M-1)\Delta f} & \ldots & e^{-j2\pi (M-1)(M-1)\Delta f}
\end{array}
\right]
\end{eqnarray}
as the DFT matrix, where $\Delta f=1/M$. 
It is obvious that $\mathbf F\mathbf F^{\mathcal H}=\mathbf I$.
As in Fig.\ref{aoa_narrow_as}, for any $\theta_{k,p}\in [-\frac{\delta\theta}{2}+\theta_k, \frac{\delta\theta}{2}+\theta_k]$, denote $\mathbf a_{k,p}=\mathbf F^{\mathcal H}\mathbf a(\theta_{k,p})$, which decomposes the steering vector
$\mathbf a(\theta_{k,p})$ using inverse DFT. From Fig. \ref{aoa_narrow_as}, only the $k_1$th to the $k_2$th elements of $\mathbf a_{k,p}$ will be significant, while other elements are negligibly small. Therefore, we have
\begin{eqnarray}
{\mathbf h}_k 
& \approx & {\tilde {\mathbf A}}_k {\tilde {\boldsymbol \alpha}}_k,
\label{eq:gao2}
\end{eqnarray}
where
\begin{eqnarray}\label{tAk}
\tilde{\mathbf {A}}_k=
\left[
\begin{array}{ccc}
 1 & \ldots & 1 \\
 e^{-j2\pi k_1 \Delta f } & \ldots & e^{-j2\pi k_2 \Delta f } \\
 \vdots & \vdots & \vdots \\
e^{-j2\pi (M-1)k_1 \Delta f} & \ldots & e^{-j2\pi (M-1)k_2 \Delta f}
\end{array}
\right],
\end{eqnarray}
\begin{eqnarray*}
\tilde {\boldsymbol \alpha}_k=\frac{1}{\sqrt{MP}}\left[
\sum_{p=1}^P a_{k,p}(k_1)\alpha_{k,p},\, \ldots,\, \sum_{p=1}^P a_{k,p}(k_2)\alpha_{k,p}
\right]^T,
\end{eqnarray*}
and $a_{k,p}(i)$ is the $i$th element of $\mathbf a_{k,p}$. 
See the derivation in Appendix B.

If $\delta\theta$ is small, then $\tilde P=k_2-k_1+1$ is also small and $\tilde{P}\ll P$.
Consequently, the decomposition of $\mathbf h_k$ in (\ref{eq:gao2}) is much more compact than that in (\ref{eq:hkmatrix}), which facilitates channel estimation and hybrid precodng design in MU massive MIMO systems. Obviously, the accuracy of the approximation in (\ref{eq:gao2}) depends on $M$, $\delta\theta$, $\tilde{P}$, $P$, as well as the instantaneous ${\boldsymbol\alpha}_k$. With (\ref{eq:gao2}), now the AOA estimation becomes to estimate $\tilde P$ equivalent AOAs or orthogonal ``frequencies," namely $k_1\Delta f,\ldots, k_2\Delta f$, which can be done by applying the $M$-point DFT on $\mathbf h_k$. In the following, we will not discriminate the term equivalent AOA and AOA, and also call estimating $k_1\Delta f,\ldots, k_2\Delta f$ as AOA estimation.

 \begin{remark}
It is also noteworthy that the above AOA estimation is performed per user even if the users have some common paths because their channels $\mathbf h_k$ can be estimated/separated first by orthogonal pilots. Hence, the subsequent AOA estimation is immune to UL IUI caused by the common paths, which is not achievable by the conventional blind AOA estimation approaches.
\end{remark}

\begin{remark}
What we have done until now is exactly the spatial sampling in the area of array signal processing, where the sampling interval is $D$ and the number of samples is $M$. Recalling $f_{k,p}=\frac{D}{\lambda}\cos(\theta_{k,p})$, we know that decreasing $D$ will not affect the range of AOA estimation.
Increasing $D$ to over $\lambda/2$, the range of $f_{k,p}$ will be beyond the capable estimation range of $[-0.5, 0.5)$.
A well-known explanation is to consider $D$ as the sampling period in the time-domain sampling\cite{vantrees}. A small $D$ will increase the sampling frequency, and therefore will not affect the estimation. On the contrary, $D$ larger than $\lambda/2$ will cause aliasing. Meanwhile, the DFT actually transfers the spatial domain to the angle/beam domain.
\end{remark}

\section{Angle Domain aided Multi-User Hybrid Preocding}

A natural question now is what can we benefit from the knowledge of the AOAs? Since the AOAs are obtained from the channels, one may expect that it can help design the  DL precoding. For hybrid transmitter where there are a limited number of RF chains, we  will show that the overall DL precoding can be decomposed into angular beamforming towards the users and the beam domain precoding that resolves the remaining IUI. Depending on whether the angular beamformings of different users are orthogonal or not, we provide the following two different designs, respectively.

\subsection{Orthogonal Beamforming and Beam Domain Precoding}
Since $f_{k,p}=\frac{D}{\lambda}\cos(\theta_{k,p})$, we may estimate the \emph{cosine} of the AOA rather than the AOA itself. Therefore, the estimation has a uniform resolution on $f$, but a non-uniform resolution on $\theta$, as we can see in Fig.\ref{aoa_narrow_as}. The equally spaced $f$ forms an orthogonal beamspace (OBS) with the corresponding steering vectors.

\begin{remark}
Another type of OBS in \cite{adhikary,yin} is obtained from the eigenvector  of the channel covariance matrix. However, such an eigenvector has no steering vector structure and the corresponding OBS is determined by both the AOAs and the path gains of the users. Obviously, eigen-space beamforming is more accurate than the angle-space beamforming but the former requires the statistic knowledge of the channel, which is hard to obtain for massive MIMO. Furthermore, eigen-space beamforming is not suitable for hybrid transmission since the eigenvector lacks the Vandermonde structure and therefore requires full digital operation on each antenna weight. Nevertheless, the angle-based OBS can  asymptotically approach the eigen-based one when the number of antennas at the BS, $M$, is very large \cite{adhikary}.
\end{remark}

From the previous section, one way to project the channels of all $K$ users onto the OBS can be done by the DFT matrix $\mathbf F$ with angular resolution $1/M$ and only a small number of orthogonal beams are significant. Let $\mathbf g_k=\mathbf F \mathbf h_k$ denote the beam domain channel of the $k$th user. If the angle spread is within a small region, then from (\ref{eq:gao2}) there is
\begin{dmath}\label{gkapprox}
\mathbf g_k \approx 
\frac{1}{\sqrt{P}}\left[\overbrace{0,\ldots,0}^{M-k_1}, \sum_{p=1}^P a_{k,p}(k_1)\alpha_{k,p},\, \ldots, 
\, \sum_{p=1}^P a_{k,p}(k_2)\alpha_{k,p}, \overbrace{0,\ldots,0}^{2k_1-k_2}\right]^T.
\end{dmath}
Nevertheless, we use
\begin{eqnarray}\label{userbeamindex}
I_k=\{i_{k,1}, \ldots, i_{k,q_k}\}
\end{eqnarray}
to represent the index set of $q_k$ elements of $\mathbf g_k$, whose magnitudes are larger than a threshold $\gamma$. Obviously, $I_k$ is a subset of $\{0, \ldots, M-1\}$.
The index of total beams occupied by $K$ users is given by
\begin{eqnarray}\label{dotGindex}
I=\displaystyle{\cup _{k=1}^K} I_k=\{i_{1}, \ldots, i_Q\}
\end{eqnarray}
whose size is denoted by $Q$.

Since the number of users is much fewer than the number of antennas at the BS, that is, $K\ll M$ and further considering the possible overlapped beams among different users, it is reasonable to assume that $Q$
is much smaller than $M$, that is $Q\ll M$.
Specifically, given the $M \times K$ channel matrix
\begin{eqnarray}\label{H}
\mathbf H = [\mathbf h_1, \ldots, \mathbf h_K],
\end{eqnarray}
and denote
$
\mathbf G = \mathbf F\mathbf H.
$
Then, only $Q$ rows $i_1,\ldots, i_Q$ in $\mathbf G$ have significant norms.
By removing $M-Q$ rows of $\mathbf G$ with small norms, we have
${\bar{\mathbf G}}\approx {\bar{\mathbf F}}\mathbf H$
or
$\mathbf H \approx \bar {\mathbf F}^{\mathcal H} \bar {\mathbf G}$,
where
$
\bar {\mathbf F}= [ \mathbf f^{\mathcal H}_{i_1}, \ldots, \mathbf f^{\mathcal H}_{i_{\dot Q}}]^{\mathcal H},
$
and $\mathbf f_m$ is the $m$th row of the $M\times M$ DFT matrix.

Owing to the channel reciprocity, the DL channel is
$\mathbf H^T \approx \bar {\mathbf G}^T \bar {\mathbf F}^*$,
where $\bar {\mathbf G}^T$ is an equivalent DL MIMO channel whose size
is $K\times Q$.
Then, the corresponding DL precoding matrix is given by
\begin{eqnarray}\label{precoder}
\mathbf P_{DL}= \mathbf B \mathbf P,
\end{eqnarray}
where $\mathbf B = {\bar {\mathbf F}}^T$ and $\mathbf P$ are the $M\times Q$ orthogonal beamforming matrix and the $Q\times K$ beam domain precoding matrix, respectively.
Because of $\bar {\mathbf F}^* \mathbf B= \mathbf I_{Q\times Q}$, as long as $Q\ge K$, $\mathbf P$ can be calculated from $\bar{\mathbf G}^T$ using the existing precoding schemes, for example, the well-known minimum-mean square error (MMSE) precoding in \cite{joham,peel}. Let the transmission power constraint be $\rho_{DL}$. The MMSE precoder can be calculated as
\begin{eqnarray}
\mathbf P=\bar{\mathbf G}(\bar{\mathbf G}^T\bar{\mathbf G}^*+\frac{K}{\rho_{DL}}\mathbf I_{K\times K})^{-1},
\end{eqnarray} and then normalized by $\sqrt{\frac{tr(\mathbf P^{\mathcal H}\mathbf P)}{\rho_{DL}}}$.

Since the beamforming vectors are orthogonal, no interference exists among formulated beams.
The IUI left is to be handled by the subsequent beam domain precoding matrix, which is much smaller in size.
Meanwhile, from (\ref{precoder}), it is clear that the OBS-based precoding is a perfect match to the hybrid precoding since $\mathbf B$ consists only phase shifters and can be implemented in the analog domain, while $\mathbf P$ is with much smaller size and can be implemented in the digital domain.
Furthermore, \emph{ the AOA estimation resolution can be directly linked to the finite resolution of phase shifter in the analog precoding}.

Let $\bar Q$ be the number of available RF chains. When $Q> \bar Q$, a beam selection process to find the best $\bar Q$ beams for transmission  is  necessary.
In practical design, $\bar Q$ is usually fixed and a smaller one is always better in terms of the cost.
It is known that the minimum number of RF chains for a $K$-user system is $K$\cite{yu}. Then we can limit the following discussion to $\bar Q=K$ due to the stringent cost requirement, for example, in a mmWave massive MIMO system.
Below, we provide a beam selection algorithm, which basically selects the most significant beam of each user by substituting $\bar Q=K$ into the function. Since the proposed algorithm is composed of $K+1$ sorting operations, the computational complexity is $\mathcal{O}((K+1)(M\log{M}))$.

 \begin{algorithm}
   \caption{Significant Beams Selection}
    \begin{algorithmic}[1]
      \Function{SigBeamSel}{$\mathbf G, \gamma, \bar Q$}
      \Comment{Where $\mathbf G$ - beam domain channel matrix, $\gamma$ - threshold, $\bar Q$ - number of available RF chains}

       \State Obtain $M$ and $K$ from the size of $\mathbf G$
       \If {$\bar Q<K$} return error \EndIf
       \For {$k = 1$ to $K$}
       \State $g_k^{max} = \max (|\mathbf g_k |)$
       \For {$m = 1$ to $M$}
      	\If {$|G(k,m)| > \gamma \frac{|| \mathbf g_k ||}{\sqrt{M}}$}
       	record $m$ to $I_k$
       	\EndIf
       	\If {$|G(k,m)| =g_k^{max}$}
       	record $m$ to $I^{max}$
       	\EndIf
       	\EndFor
       	\EndFor
       	\State $I=\cup_{k=1}^K I_k$ and record its size $Q$
       	\State Form the matrix $\bar {\mathbf G}$ from $\mathbf G$ using $I$
       	\State Sort the rows of $\bar {\mathbf G}$ in descending order of norm
       	\State Reorder $I$ accordingly
       	\If {$Q > \bar Q$}
       	\State Let $\vec I$ be the first $\bar Q-K$ indexes of $I \setminus I^{max}$
       	\State return $I^{max} \cup \vec I$
       	\Else
       	\State return $I$      	
       	\EndIf

       \EndFunction

\end{algorithmic}
\end{algorithm}

\subsection{Non-Orthogonal Beamforming and Beam Domain Precoding}

The OBS based scheme relies on the fixed-directional beams towards the equally spaced AOAs due to the usage of the $M\times M$ DFT matrix. Hence, OBS has two critical drawbacks:
\begin{itemize}
\item[(i)] The AOA resolution is only $1/M$;
\item[(ii)] The orthogonal beams obtained may not necessarily point to the strongest direction of users, and thus will suffer from \emph{power leakage}.
\end{itemize}

If the  AoA of the user is not exactly an integer times of $1/M$, then the DFT leakage will cause wider beam occupancy and subsequently a large $Q$. In other words, the orthogonal beamforming constraint may bring unnecessary beam spread.  Since we only have $K$ RF chains, that is, one beam for each user, it is desirable to
improve the accuracy of the analog beamforming such that the beam domain channel gain of a single element can be maximized. An effective way is to consider non-orthogonal beams to suppress the DFT leakage and narrow the beam spread. In fact, the non-orthogonal beams can be easily obtained via rotating the beam by a small angle such that the beams will point to the strongest direction of the users.

To rotate OBS, let us introduce a refining angle for each user. Denote $\psi_k$ to be the refining angle for user $k$, and
\begin{eqnarray}\label{iar}
\frac{-1}{2M}\le \psi_k \le \frac{1}{2M}, 1\le k \le K,
\end{eqnarray}
we can also denote
\begin{eqnarray}
\boldsymbol O(\psi_k) = {\rm diag} \{ 1, e^{j2\pi\psi_k}, \ldots, e^{j2\pi(M-1)\psi_k} \}.
\end{eqnarray}
Similar to (\ref{gkapprox}), we can calculate
\begin{eqnarray}
\tilde{\mathbf g}_k(\psi_k)=\mathbf F \boldsymbol O(\psi_k)\mathbf h_k.
\end{eqnarray}
Then, the optimal refining angle for the $k$th user will be determined by
\begin{eqnarray}\label{bia}
\psi_k^0= \argmax_{\psi_k}\  {\tilde g}_k^{max}(\psi_k),
\end{eqnarray}
where ${\tilde g}_k^{max}(\psi_k)$ is the maximum element of $\tilde {\mathbf g}_k(\psi_k)$ and the position of ${\tilde g}_k^{max}(\psi_k)$ in $\tilde {\mathbf g}_k(\psi_k)$ is denoted as $\tilde i_k$.

\begin{figure}[t]
\centering
\includegraphics[width=9cm]{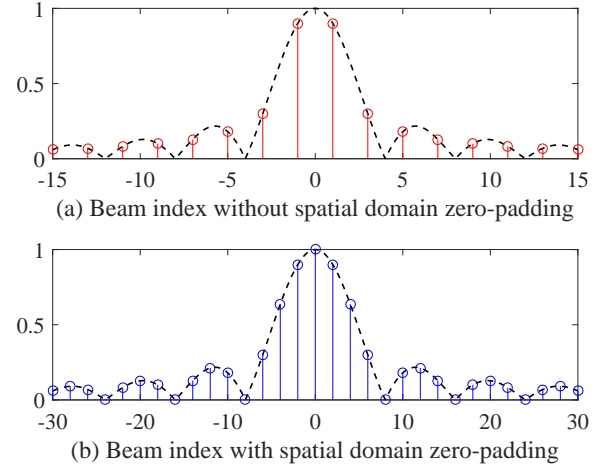}
\caption{Beam domain oversampling by zero-padding.}
\label{beamsinc}
\end{figure}

The estimation of the optimal refining angle in (\ref{bia}) needs a one-dimensional search over the range in (\ref{iar}). However, bearing in mind that $\psi$ is a small $f$, we actually need a high resolution ``frequency"  estimation via digital approach with finite grid. With the DFT operation, such high resolution estimation can be performed by padding zeros  at the end of channel vector.
Let $V$ be an integer lager than $1$, we can calculate
\begin{eqnarray}\label{GVM}
\mathbf g_{k,VM} = \sqrt{V} \mathbf F_{V M}
\left[
\begin{array}{c}
\mathbf h_k  \\
\mathbf 0_{(V-1)M\times 1}
\end{array}
\right],
\end{eqnarray}
where $\mathbf F_{V M}$ is the $VM\times VM$ DFT matrix.
Then, we can obtain $V$ vectors $\mathbf g_{v,k}, 0 \le v \le V-1$
where the $v$th vector $\mathbf g_{v,k}$ is formed by the $(v+mV)$th entry of $\mathbf g_{k,VM}$ with $0 \le m \le M-1$.
From the basic sampling theory, we know
\begin{eqnarray}
\mathbf g_{v,k} = \tilde {\mathbf g}_k\left(\frac{v}{VM}\right)
\end{eqnarray}
which implies that the cost function  in (\ref{bia}) is evaluated for $\psi_k=\frac{v}{VM}, 0 \le v \le V-1$.
Then, one can easily obtain $\psi_k^0$  and $\tilde i_k$ by comparing the $VM$ elements in $\mathbf g_{v,k}$, $0 \le v \le V-1$.
Compared to the original $\mathbf{g}_k$, the beam domain resolution improves from $1/M$ to $1/VM$, as illustrated in Fig. \ref{beamsinc} with $V=2$ for a simple case of $P=1$.

Hence, a valid approximation to $\mathbf{h}_k$ with one beam vector is written as
\begin{equation}
\mathbf{h}_k=\mathbf O^{*}(\psi_k^0)\mathbf f^{\mathcal H}_{i_k} {\tilde g}_k^{max}(\psi_k^0) \label{eq:gao3}
\end{equation}
where $\mathbf f_{\tilde i_k}$ is the $\tilde i_k$th row of $\mathbf{F}$. In fact, steps (\ref{GVM}) to (\ref{eq:gao3}) can be viewed as high-resolution AOA  estimation for the $k$th user.

\begin{figure}[t]
\centering
\includegraphics[width=8cm]{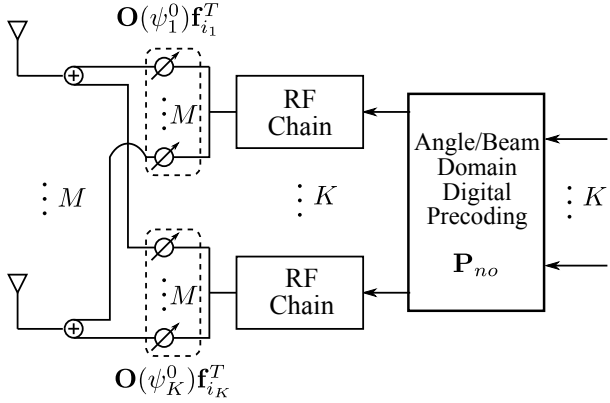}
\caption{NOAS-based hybrid precoding.}
\label{noas}
\end{figure}

\begin{remark}
 Since we have $M$ spatial samples of the user channel, the $M$-point DFT
only provides $M$ angle/beam domain responses within a ``bandwidth" of $1/\lambda$. The zero-padding and the $VM$-point DFT actually further divide the ``bandwidth" by $VM$ to obtain $VM$ angle/beam domain responses.
\end{remark}

The overall channel matrix for all $K$ users can be approximated by
\begin{eqnarray}
\mathbb H \approx [\mathbf O^{*}(\psi_1^0)\mathbf f^{\mathcal H}_{i_1} {\tilde g}_1^{max}(\psi_1^0), \ldots,  \mathbf O^{*}(\psi_K^0)\mathbf f^{\mathcal H}_{i_K} {\tilde g}_K^{max}(\psi_K^0)],
\end{eqnarray}
for which  the strongest beam of each user channel is chosen and is handled with only one RF chain.
Hence, the non-orthogonal analog beamforming vector for the $k$th user can be selected as $(\mathbf O^{*}(\psi_k^0)\mathbf f^{\mathcal H}_{i_k} )^{*}= {\mathbf O}(\psi_k^0)\mathbf f^T_{i_k}$. Bearing in mind that $\mathbf O(\psi_k^0)$ is a diagonal matrix, the analog precoding matrix becomes
\begin{eqnarray}
\mathbf B_{no} = [\mathbf O(\psi_1^0)\mathbf f_{i_1},  \ldots, \mathbf O(\psi_K^0)\mathbf f_{i_K}]^T.
\end{eqnarray}
The $K$ non-orthogonal beams $\mathbf O(\psi_k^0)\mathbf f_{i_k}$ actually form a non-orthogonal beamspace, or strictly speaking, non-orthogonal angle space (NOAS), pointing to the strongest direction of each user while the caused inter beam interference (IBI) will be handled by the digital precoding part.\footnote{Note that, the IUI is made up of the IBI among analog beamformers and the remained IUI in the angle/beam domain.} With the NOAS-based beamforming, the precoding matrix $\mathbf P_{no}$ can be calculated from $\bar {\mathbf G}^T\bar {\mathbf F}^* \mathbf B_{no}$, where the calculation of $\bar {\mathbf F}^* \mathbf B_{no}$ can be significantly simplified using the result in (\ref{ckn}).

The structure of the proposed NOAS-based hybrid precoding schemes is shown in Fig. \ref{noas}. In practice, the phase shifter with continuous variable phase is not only inaccurate but also expensive. In contrast, a phase shifter with finite phase shift is with low-cost and can be controlled precisely.
In the OBS-based hybrid precoding ($V=1$), we basically need $M\times K$ phase shifters with a relatively-lower resolution of $1/M$.
The high-resolution NOAS-based hybrid precoding ($V>1$) requires $M\times K$ phase shifters with a high resolution of $1/(VM)$.
Therefore, only finite resolution phase shifters are required in the proposed hybrid precoding scheme and a low-cost implementation is possible.

\subsection{Summary of Different Beamforming Methods  in Massive MIMO}

\begin{figure} \centering
  \vspace{-10pt}
 \subfigure[Eigen-space beamforming] { \label{fig:1a}
 \includegraphics[width=8cm]{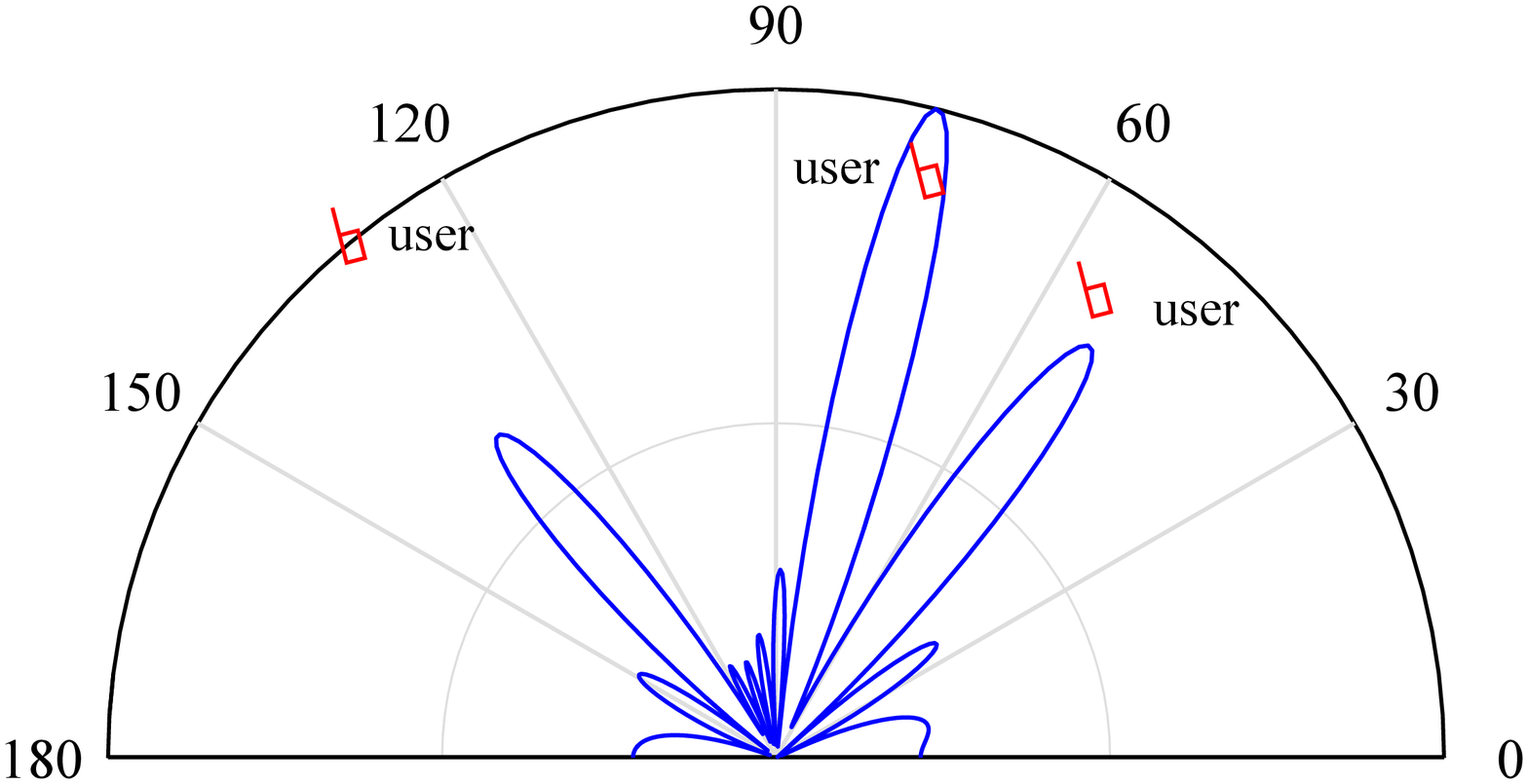}
 }

 \subfigure[Orthogonal angle space (beamspace) beamforming] { \label{fig:1b}
 \includegraphics[width=8cm]{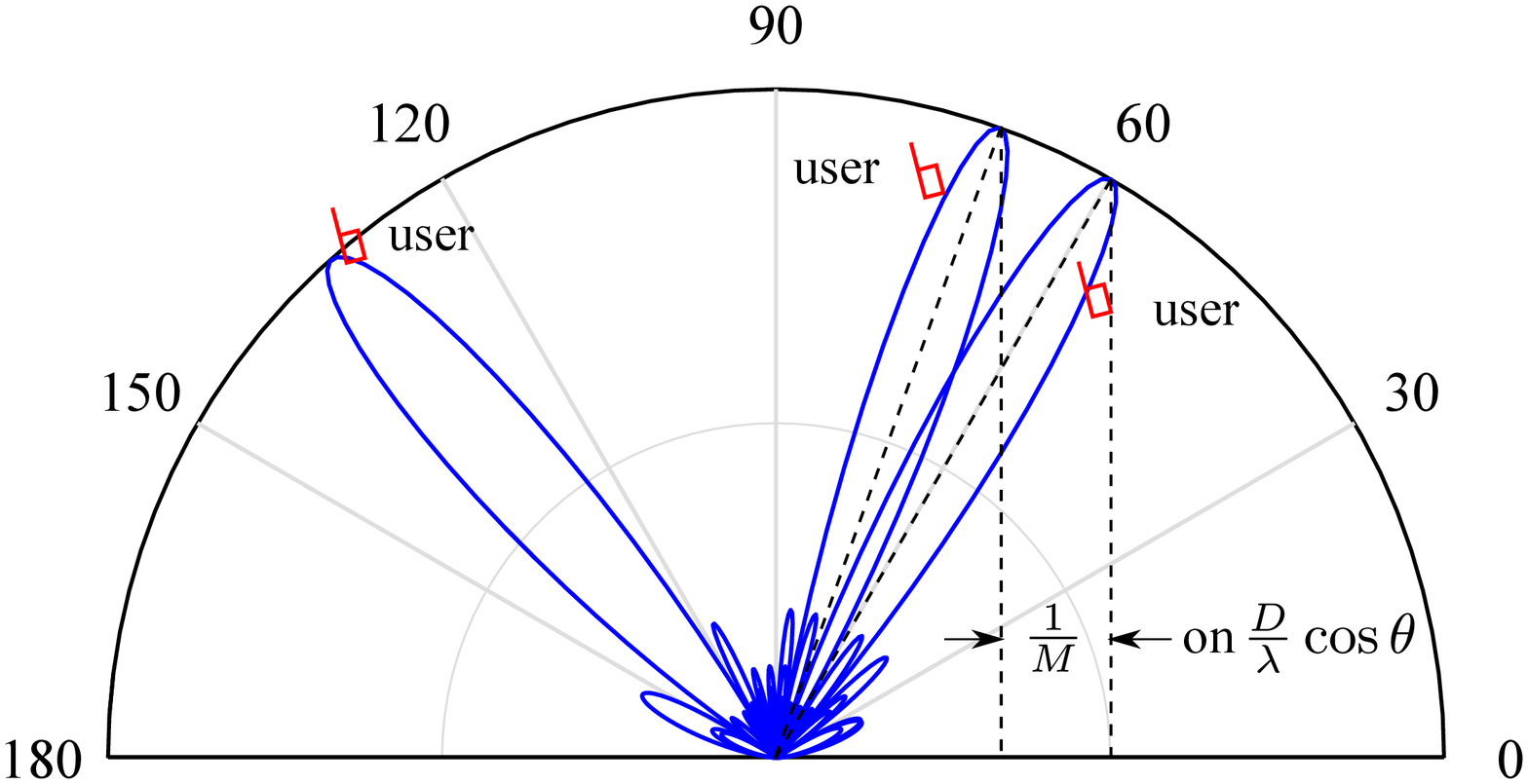}
 }
  \subfigure[Non-orthogonal angle space beamforming ] { \label{fig:1c}
 \includegraphics[width=8cm]{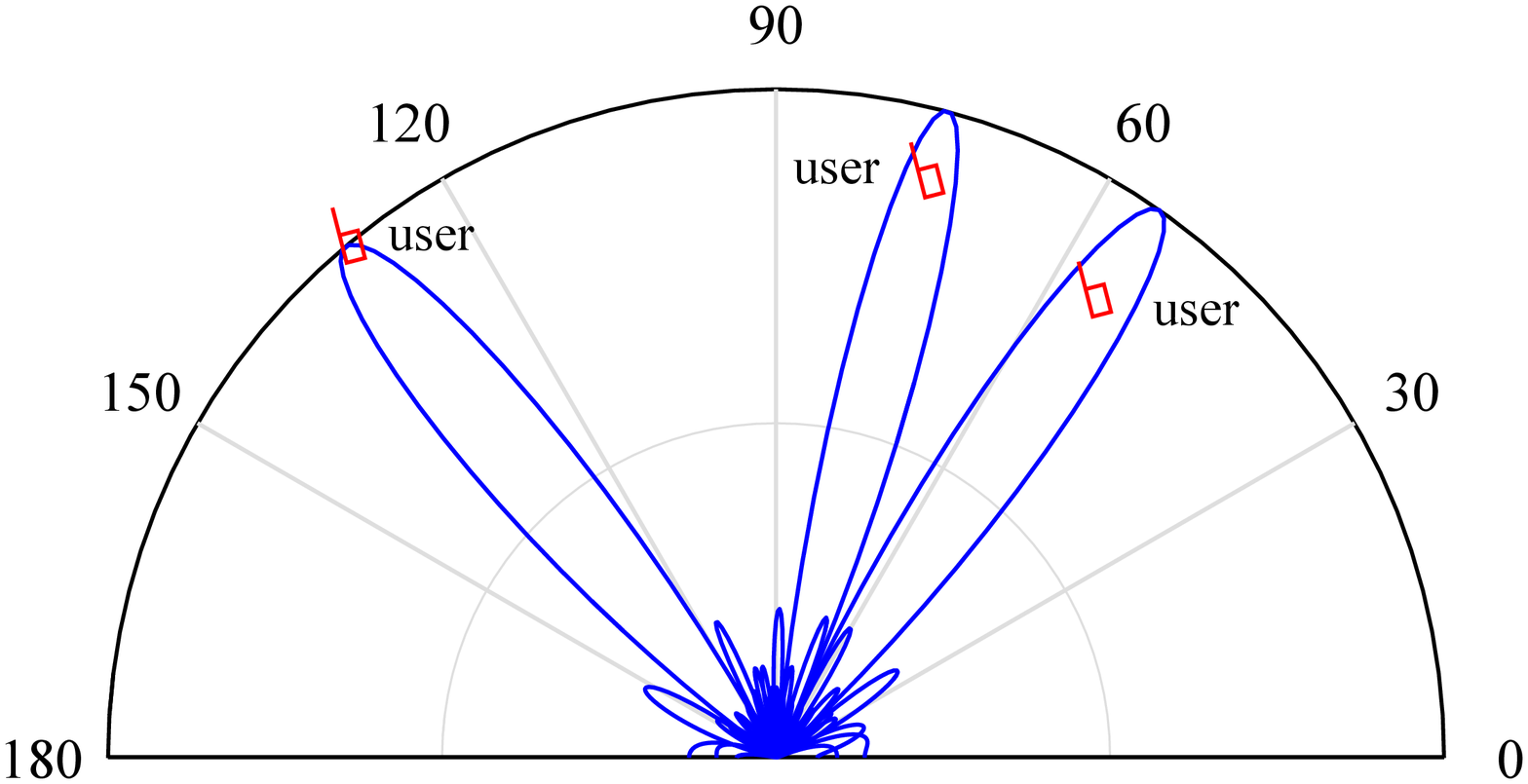}
 }
 \caption{ Illustration of three different beamforming methods. }
 \label{fig:gao1}
\end{figure}

Till now, there are three beamforming methods in the literature from  different space viewpoint, which are summarized as follows:
\begin{enumerate}
\item As in Fig. \ref{fig:1a}, the eigen-space method utilizes the best eigen-directions for beamforming \cite{adhikary}, \cite{yin}.
Specifically, the beamforming vectors of this method do not physically formulate beams towards users but
rather change the amplitude and the phase of each antennas such that the optimal signal-to-interference-noise (SINR) ratio (or other criterion) can be  achieved at users. Hence, the method is only valid for full digital operation.

\item As in Fig. \ref{fig:1b}, the angle-space method utilizes the orthogonal steering vectors for beamforming, as given in Section IV.A and  in \cite{cycling}, \cite{Sayeed} (sometimes called beamspace method). There are only $M$ such steering vectors that formulate orthogonal beams towards the fixed directions (codebook based beamforming).  Hence, the beampspace method can be viewed as ``angle-on-grid"  method but does not identify the ``true angle'' of the users. Since the beam directions generally do not point to the exact directions of users, the beamspace method will suffer from power leakage \cite{sbem,sbem1,sbem2}. This method is  valid both for full digital and hybrid operation.

\item As in Fig. \ref{fig:1c}, the angle-space method utilizes the non-orthogonal steering vectors for beamforming, as presented in Section IV.B. Specifically, the beamforming vectors of this method are also chosen from the steering vectors but the beams  point to the exact directions of users (user-centric beamforming). Hence, the  beams are non-orthogonal to each other and there will be IBI. This method is valid both for full digital and hybrid operation.

\end{enumerate}

\subsection{ Angle Division Multiple Access  }

The observation from the spatial and the angle/beam domains implies that in the case of a single user, the ULA-based massive MIMO system can be viewed as an orthogonal beam division multiplexing (OBDM) system (with more than one RF chains for this user from orthogonal columns of DFT matrix), which is analogous to the orthogonal frequency division multiplexing (OFDM) system as compared in Table~I. It is then clear that the existing research results of OFDM systems can be directly applied to the single SU massive MIMO system by projecting the user channel onto the DFT-based OBS.

On the other hand, for the case of MU multiple access, there is a fundamental difference between the massive MIMO and the orthogonal frequency division multiple access (OFDMA) systems. In OFDMA systems, the bandwidth is defined on the frequency domain, and therefore  the orthogonality among different users can be easily maintained. In contrast,  the ``bandwidth'' in the massive MIMO systems is defined on the angle/beam domain and depends on the users spatial location. Therefore the orthogonality among different users is out of our control. Consequently, an MU massive MIMO system could either be an orthogonal ADMA system by using OBS or be a non-orthogonal ADMA system by using NOAS, where the former removes IBI by sacrificing on power leakage while the latter does reversely. It can then be imagined that when the number of the RF chains is very small, the power leakage will be the dominant issue and the NOAS would performs better. Whereas  when the number of the RF chains is very large, the power leakage is very small and the OBS may perform better.

\section{Channel and AOA Estimation with Limited RF Chains}
All previous discussions are based on the availability of channel matrix $\mathbf H$ and the corresponding AOAs. With limited RF chains, such channel estimation is normally obtained by sequentially sending the pilot
and sweeping the beam on all directions, named as beam cycling  \cite{cycling}.
However, from the angle domain viewpoint, it is possible
to reduce the amount of beam sweeping, and thus greatly reduce the training overhead.

Let $\mathbf X$ denote the $K\times K$ unitary pilot matrix in the UL. The received pilot signal at the BS is
\begin{eqnarray}
\mathbf Y= \mathbf H \mathbf X,
\end{eqnarray}
where the additive white Gaussian noise has been ignored for the sake of convenience.
Let us consider the phase shifters at the RF chains as an analog  combiner  represented by a $K\times M$ matrix $\mathbf E$. Then, we right-multiply the output of the $K$ RF chains by $\mathbf X^{\mathcal H}$ and left-multiply by $\mathbf E$ to obtain a $K\times K$ matrix
\begin{eqnarray}\label{partH}
\bar {\mathbf H}=\mathbf E\mathbf Y \mathbf X^{\mathcal H} = \mathbf E \mathbf H,
\end{eqnarray}
where we have used the property of $\mathbf X\mathbf X^{\mathcal H}=\mathbf I$. We next explore the sparsity of $\mathbf H$ in the angle domain to recover ${\mathbf H}$ from $\bar {\mathbf H}$.

The similarity of ADMA to the OFDM/OFDMA system inspires us that we are actually dealing a problem similar to frequency domain OFDMA channel estimation. In particular, we know that the user channel is quite sparse in frequency domain even if we do not know the indices of valid frequency bin.
Moreover, due to the limited number of RF chains, we only have part of its spatial (time) domain impulse response depending on how the RF chains are connected to antennas. The problem becomes \emph{OFDM sparse frequency domain channel estimation with insufficient impulse response }.

\begin{table}[t]
\centering
\caption{OBDM versus OFDM}
\begin{tabular}{|c|c|}
\hline
OBDM & OFDM  \\
\hline
spatial & time \\
\hline
angle/beam  & frequency \\
\hline
number of beams  & number of subcarriers \\
\hline
\end{tabular}
\end{table}

Due to the orthogonal pilots, these portions of user channels have been separated, namely,  the $k$th column of  $\bar {\mathbf H}$ can be expressed as
\begin{eqnarray}\label{parth}
\bar {\mathbf h}_k=\mathbf E \mathbf h_k.
\end{eqnarray}
Since we do not know the AOA information of the users, we require the training  $\mathbf X$ to be sent twice.

\subsection{First Time training: Achieve the Rough AOA Information}

If   the index of significant beams in (\ref{dotGindex}) is known, i.e., the AOA information, then we can let the phase shifters at each row of the analog combiner $\mathbf E$ act as the corresponding rows of the DFT matrix to obtain $\bar {\mathbf H}$, which will become a fine estimate of the angle/beam domain response of the spatial channel $\mathbf H$.  In other words, each row of the analog combiner can become a \emph{receive beamformer} if the AOA information is available. For the current case of $K$ RF chains (one for each user),  we actually need to know the index of the most significant beam (IMSB) for each user .

\begin{figure}[t]
\centering
\includegraphics[width=8cm]{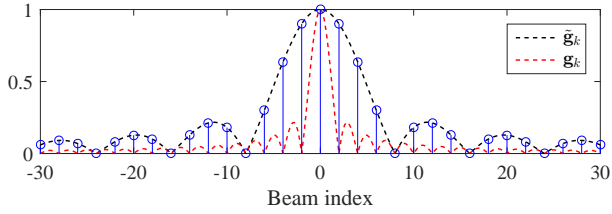}
\caption{IMSB of the $k$th user. }
\label{zp}
\end{figure}

Hence, in the first time training, we target at roughly achieving the AOA information of each user. Without loss of generality, we propose the following structure for the analog combiner matrix , 
\begin{eqnarray}
\mathbf E_0= [\mathbf I_{K\times K}, \mathbf 0_{K\times (M-K)}],
\end{eqnarray}
that is, we simply connect $K$ RF chains to the first $K$ antennas.
Then, from the previously discussions, we can apply the  DFT  approach to achieve AOA estimation from
\begin{eqnarray}\label{tildeG}
\tilde {\mathbf g}_k = \sqrt{V}\mathbf F_{MV}
\left[
\begin{array}{c}
\mathbf E_0 \mathbf h_k \\
\mathbf 0_{(VM-K)\times 1}
\end{array}
\right].
\end{eqnarray}
Since $\tilde{\mathbf g}_k$ is the DFT of an deficient sampling of $\mathbf h_k$,  $\tilde{\mathbf g}_k$  does not have the same envelope as  $\mathbf g_k$ but is rather a ``fatter version'' of $\mathbf g_k$. Hence, $\tilde{\mathbf g}_k$ would have the its maximum value at a position close to that of ${\mathbf g}_k$.
 An example of $\mathbf g_k$ and $\tilde {\mathbf g}_k$ calculated from a normalized $\mathbf h_k$ is given in Fig.~\ref{zp} with $M=64$, $K=16$, and $P=1$.
From the figure, $\tilde{\mathbf g}_k$ and ${\mathbf g}_k$, though with different shape, would have the same maximum value at the same position. Therefore, we can obtain an estimate of IMSB from  $\tilde{\mathbf g}_k$ in the first place.

\subsection{Second Time Training: Channel Estimation over the Known AOA Directions}
When $V=1$, i.e., the OBS scheme, let $\hat i_k$ be the IMSB of the $k$th user. The analog combiner $\mathbf E_1$ at the second time training is set as
\begin{eqnarray}
\mathbf E_1=[\mathbf f_{\hat i_1}^T,\ldots,\mathbf f_{\hat i_K}^T]^T.
\end{eqnarray}
With $\mathbf E_1$, which is exactly an OBS-based receive beamformer, we can obtain
\begin{eqnarray}
\bar {\mathbf H}=\mathbf E_1\mathbf Y \mathbf X^{\mathcal H}=\mathbf E_1 \mathbf H = \tilde { \mathbf G}.
\end{eqnarray}
In fact, $\tilde { \mathbf G}$ can be considered as a size-reduced approximation of the beam domain equivalent MIMO channel $\bar {\mathbf G}$.
For $V>1$, i.e., NOAS scheme, we can obtain an approximation of $\bar {\mathbf G}^T\bar {\mathbf F}^* \mathbf B_{no}$ because $\mathbf E_1$ becomes the NOAS-based receive beamformer.

\begin{figure}%
    \centering
    \subfigure[First step: IMSB estimation.]{
        \includegraphics[width=7.9cm]{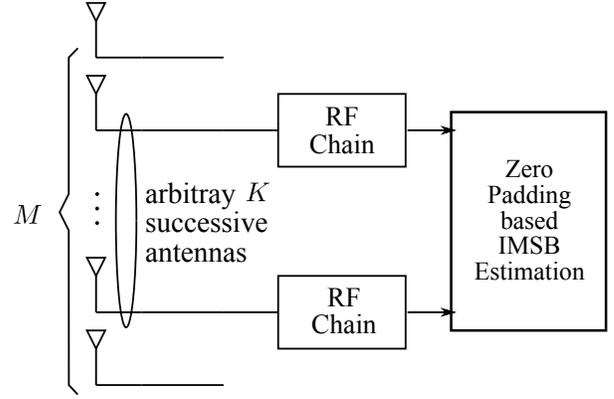}
        \label{fig_bdoe}} \hfill\\

    \subfigure[Second step: Angle domain channel estimation, $V>1$.]{
        \includegraphics[width=7.6cm]{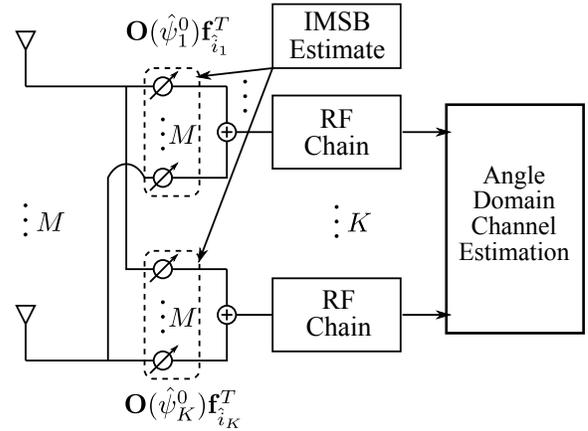}
        \label{fig_bdce} }
\caption{Two-Step channel estimation}\label{fig:two-step-ce}
\end{figure}

In summary, the proposed channel estimation approach includes two steps, the IMSB estimation and the angle domain channel estimation ($V>1$), which are shown in Fig. \ref{fig_bdoe} and Fig. \ref{fig_bdce}, respectively. Since the DL transmitting and the UL receiving share the same beamforming vectors, once the analog receive beamformer $\mathbf E_1$ is fixed, the DL transmission is just the calculation using the digital precoding matrices, corresponding to the OBS or the NOAS based beamformers, respectively.

\begin{remark}
For conventional beam cycling based hybrid channel estimation \cite{cycling}, it  needs to sweep the beam over all $M$ antennas, which costs
 $M/K$ times of training. In comparison, the proposed method needs two times of training, which greatly saves overhead.
\end{remark}

\begin{figure*}%
    \centering
    \subfigure[$K=4$, $\delta\theta=1^{\circ}$.]{
        \includegraphics[width=8.5cm]{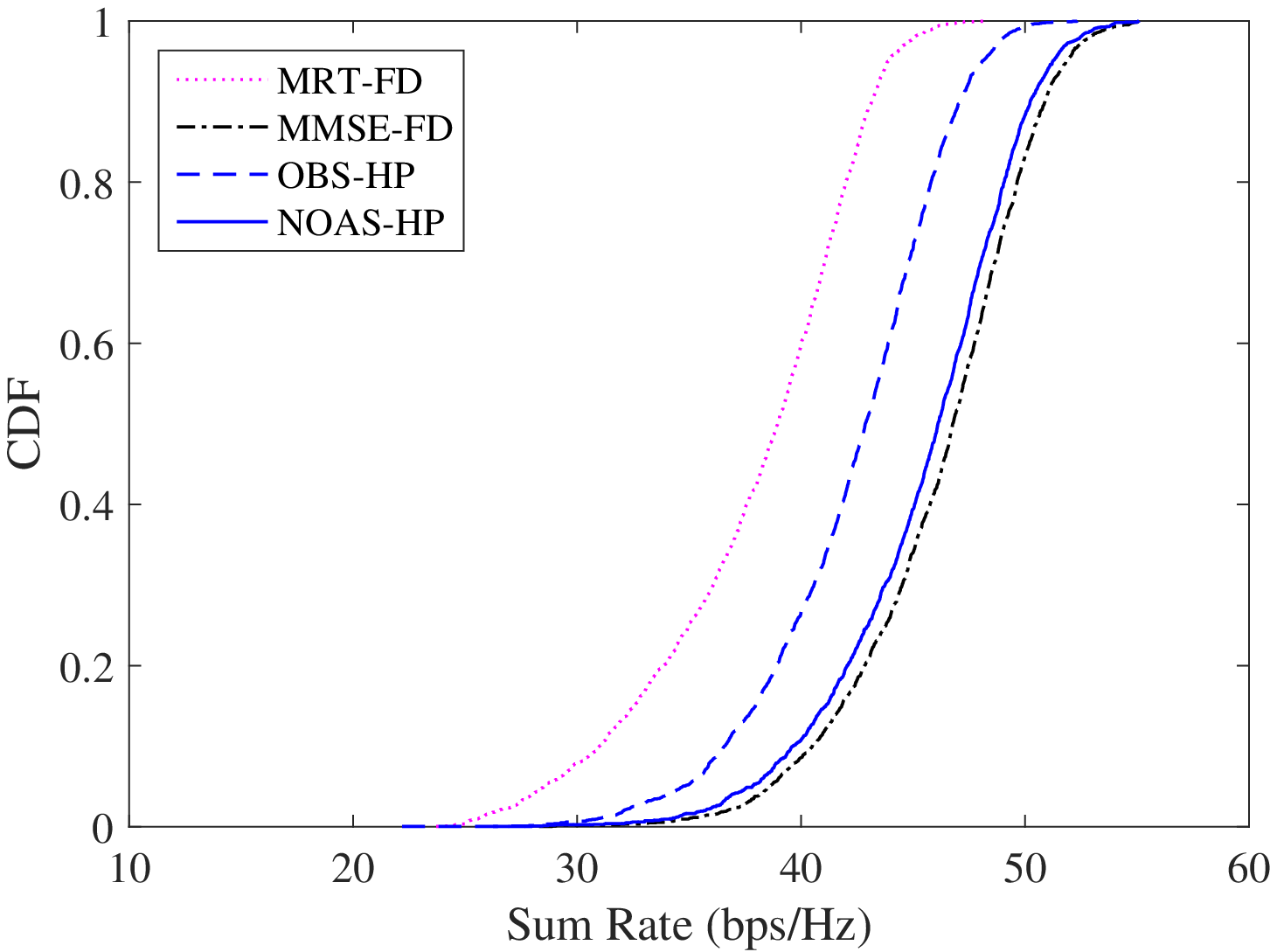}
        \label{cdf1}}
    ~%
    \subfigure[$K=8$, $\delta\theta=1^{\circ}$.]{
        \includegraphics[width=8.5cm]{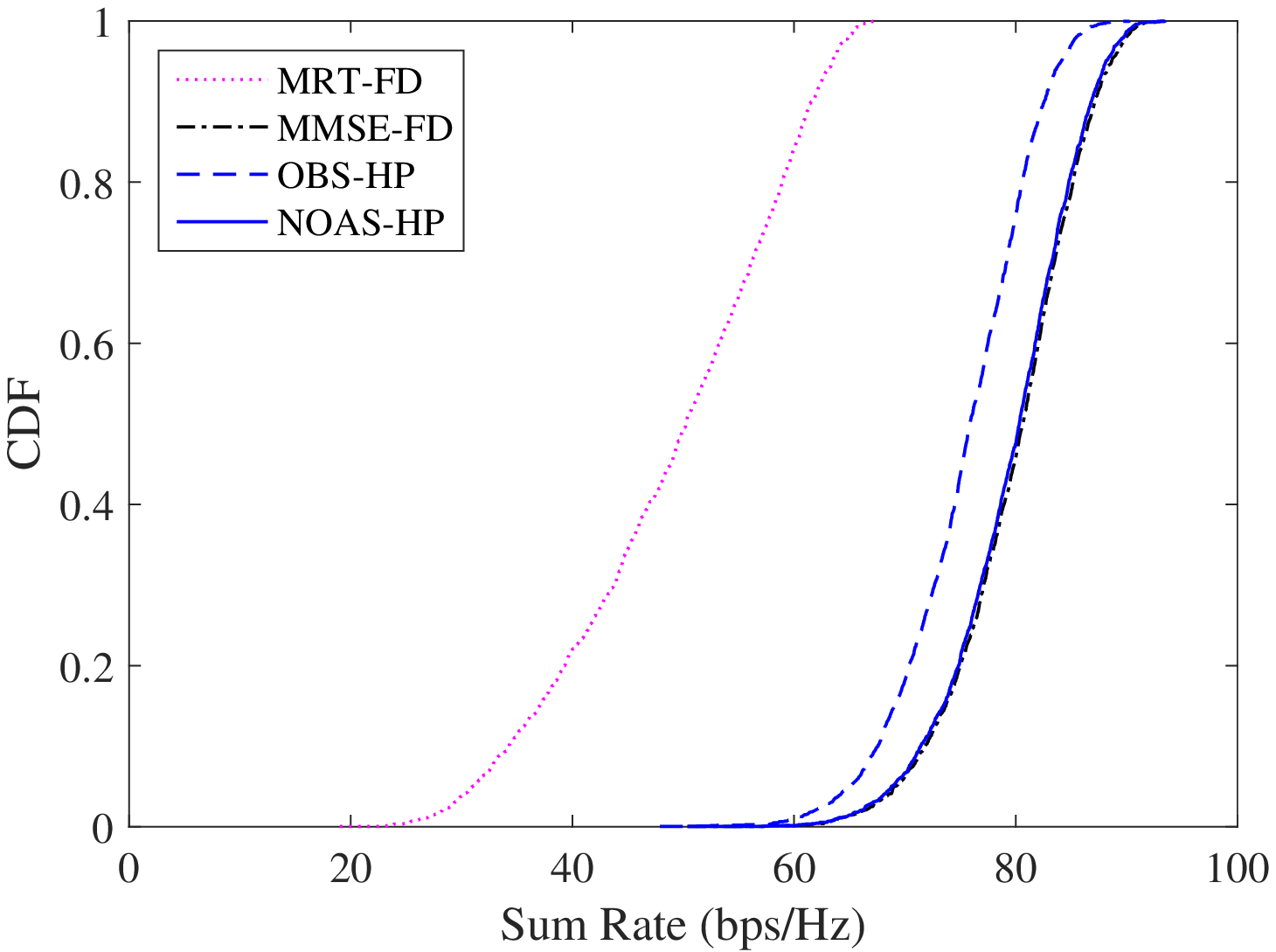}
        \label{cdf2} } \\
    \subfigure[$K=4$, $\delta\theta=3^{\circ}$.]{
        \includegraphics[width=8.5cm]{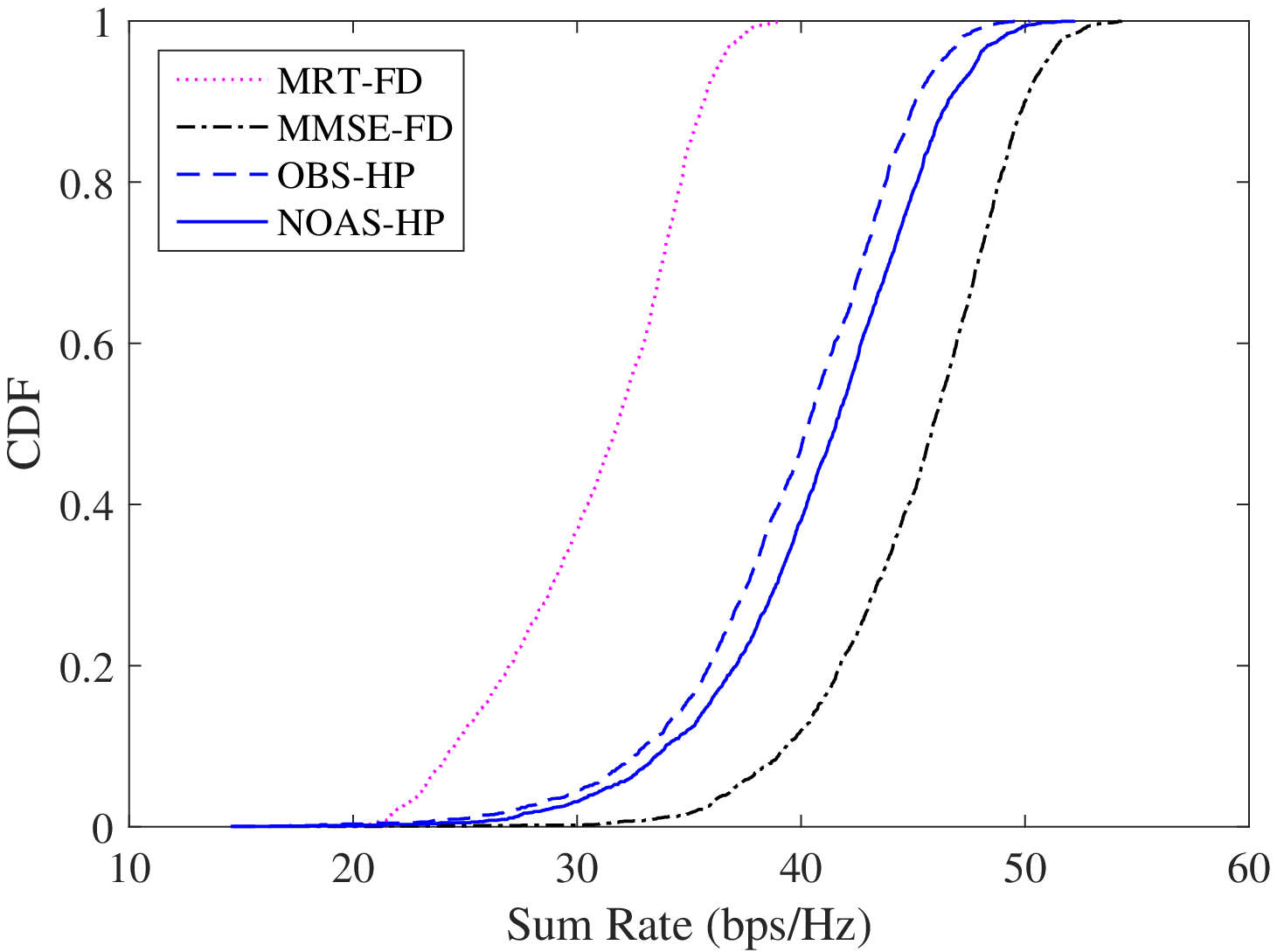}
        \label{cdf3}}
    ~%
    \subfigure[$K=8$, $\delta\theta=3^{\circ}$.]{
        \includegraphics[width=8.5cm]{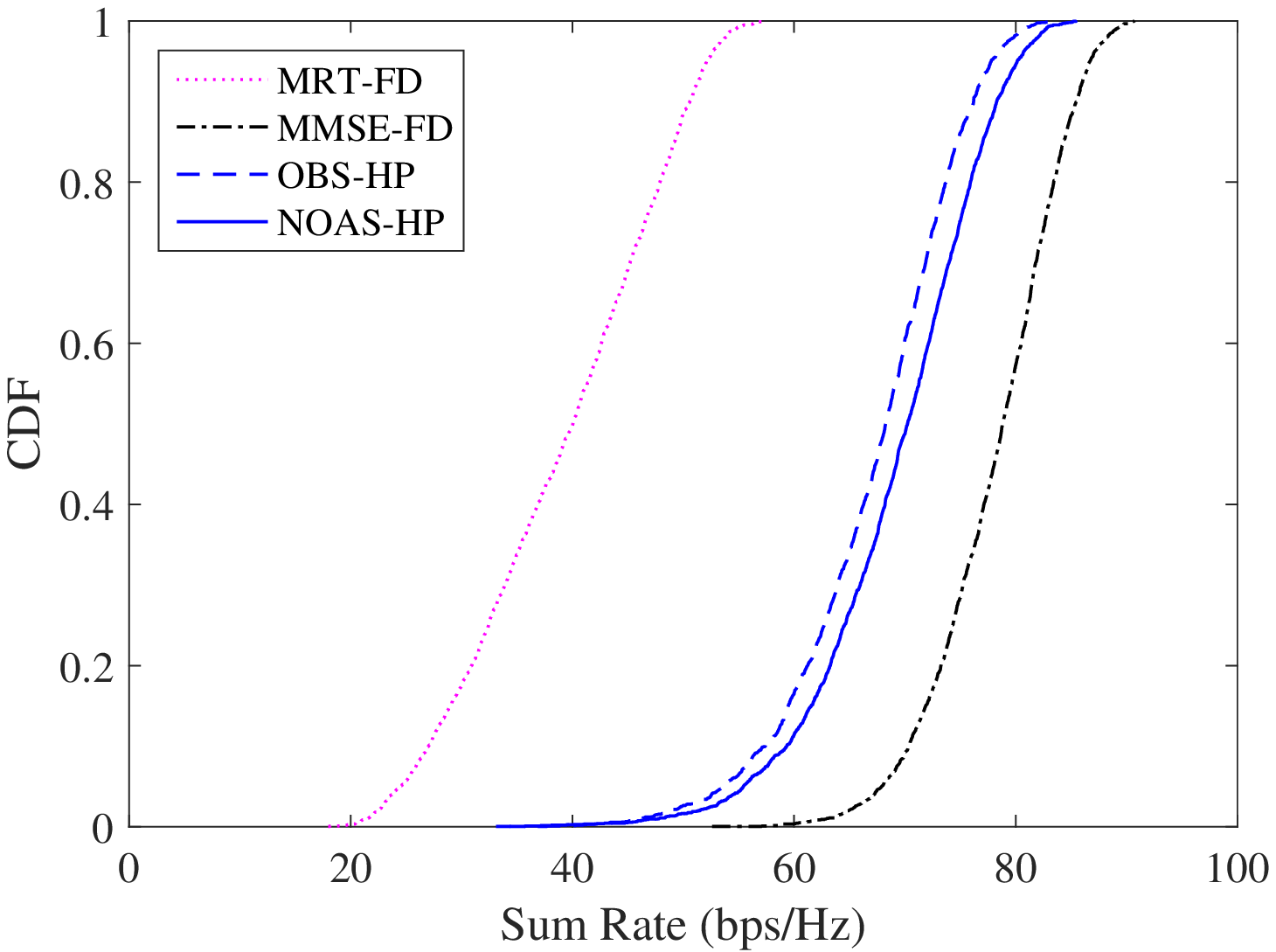}
        \label{cdf4}}
    \caption{CDF of Sum-Rate}\label{fig:cdf}
\end{figure*}

\begin{remark}
Theoretical analysis and field tests have shown that the angle of departure (AOD) of the DL channel is the same as  the AOD of the UL channel \cite{sbem}, \cite{reciprocity}, even for frequency division duplex (FDD) systems. In other words, even for FDD systems without channel reciprocity, the UL and DL channels do have angle reciprocity. Hence, the proposed angle domain channel estimation principle can also be applied to simplify the DL channel estimation in FDD system, where only a few additional pilots are needed to estimate the channel gains.
\end{remark}

\section{Simulation results}
In this section, we will present simulation results to verify our discussion before.
We consider a TDD massive MIMO system, where the ULA at the BS has $M=64$ antennas. There are $K$ single-antenna users uniformly located inside a semicircular cell with a radius of one kilometer, and the users angular spread is $\delta \theta$.
The channel fading coefficients are generated from the urban micro model in the 3GPP standard \cite{3gpp} with $P=20$. The path loss is given by
$
PL(dB)  = -35.4+26\log_{10}(d) +20\log_{10}(f_c),
$
where $d$ is the distance between a user to the BS, and $f_c=3.7$GHz. Also, a bulk log normal shadowing with a standard deviation of $4$ dB is applied to all sub-paths.  For the UL training, the signal-to-noise ratio (SNR) observed at the BS is $25$ dB. For the DL precoding, the BS power constraint is $\rho_{DL}=50$ dBm, and the noise variance at the user side is $\delta_k^2 =-92$ dBm.

The proposed NOAS-based hybrid precoding (NOAS-HP) and the OBS-based hybrid precoding (OBS-HP)\footnote{The OBS-HP coincides with the exiting beamspace method \cite{cycling}, \cite{Sayeed}.} are compared with the matched filter based maximum ratio transmission (MRT) precoding and the MMSE precoding in terms of sum-rate. The MRT and the MMSE precodings are full-digital (FD) while the digital part in the proposed hybrid precoding schemes is an MMSE precoder.

The cumulative distribution functions (CDF) of the sum rate of all $K$ users are shown in Fig.~\ref{fig:cdf} for different $K$ and $\delta\theta$, respectively.  From Fig. \ref{cdf1} and Fig. \ref{cdf2}, it  can be seen that when the angular spread is $1^\circ$, both the proposed OBS- and NOAS-based hybrid precoding schemes combined with the proposed channel estimation significantly outperform the MRT. The reason
is that when $M$ is not infinitely large and when the channel paths are not infinitely rich ($P=20$), then the real channel $\mathbf{h}_k$'s have very poor orthogonality. However, our angle domain approaches do not require such property  and hence the performances are satisfactory. Especially, the NOAS-based hybrid precoding scheme presents performance very close to the high-complexity all-digital MMSE precoding because it forms more ``focusing'' beams to increase the gain of the equivalent MIMO channel. For the same reason, the NOAS-based scheme has the sum rate $5$ bps/Hz higher than that of the OBS-HP.
With larger angular spread of $3^\circ$, the results in Fig. \ref{cdf3} and Fig. \ref{cdf4} show that MRT is still not good while MMSE becomes apparently better than the proposed methods. The reason is that the proposed methods assign only $K$ RF chains for $K$ users, i.e., one RF chain per user. In this case, the one beam approximation of the whole channel would become  worse as $\delta\theta$ increases. Nevertheless, such performance degradation could be mitigated if more RF chains are available for transmission, say in full digital transmission. Moreover, the performance gap between the OBS-HP and the NOAS-HP narrows down to around $2$ bps/Hz because a single rotation
would not catch more power of the channel with a single beam compared with non-rotation case under  wide angular spread.
Hence, the proposed hybrid schemes, especially the NOAS one, are more suitable for narrow angular spread scenario, such as in the mmWave case.

\begin{figure}      
\centering
\includegraphics[width=8.5cm]{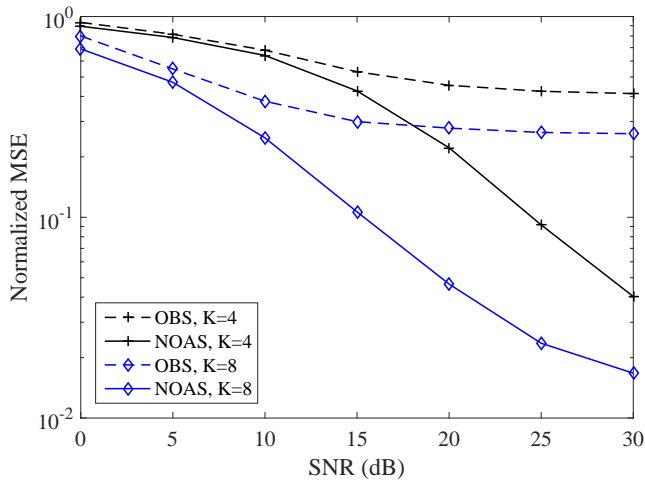}
\caption{{Normalized MSE versus UL SNR, $\delta \theta=1^{\circ}$.} }
\label{nmseA1}
\end{figure}

\begin{figure}      
\centering
\includegraphics[width=8.5cm]{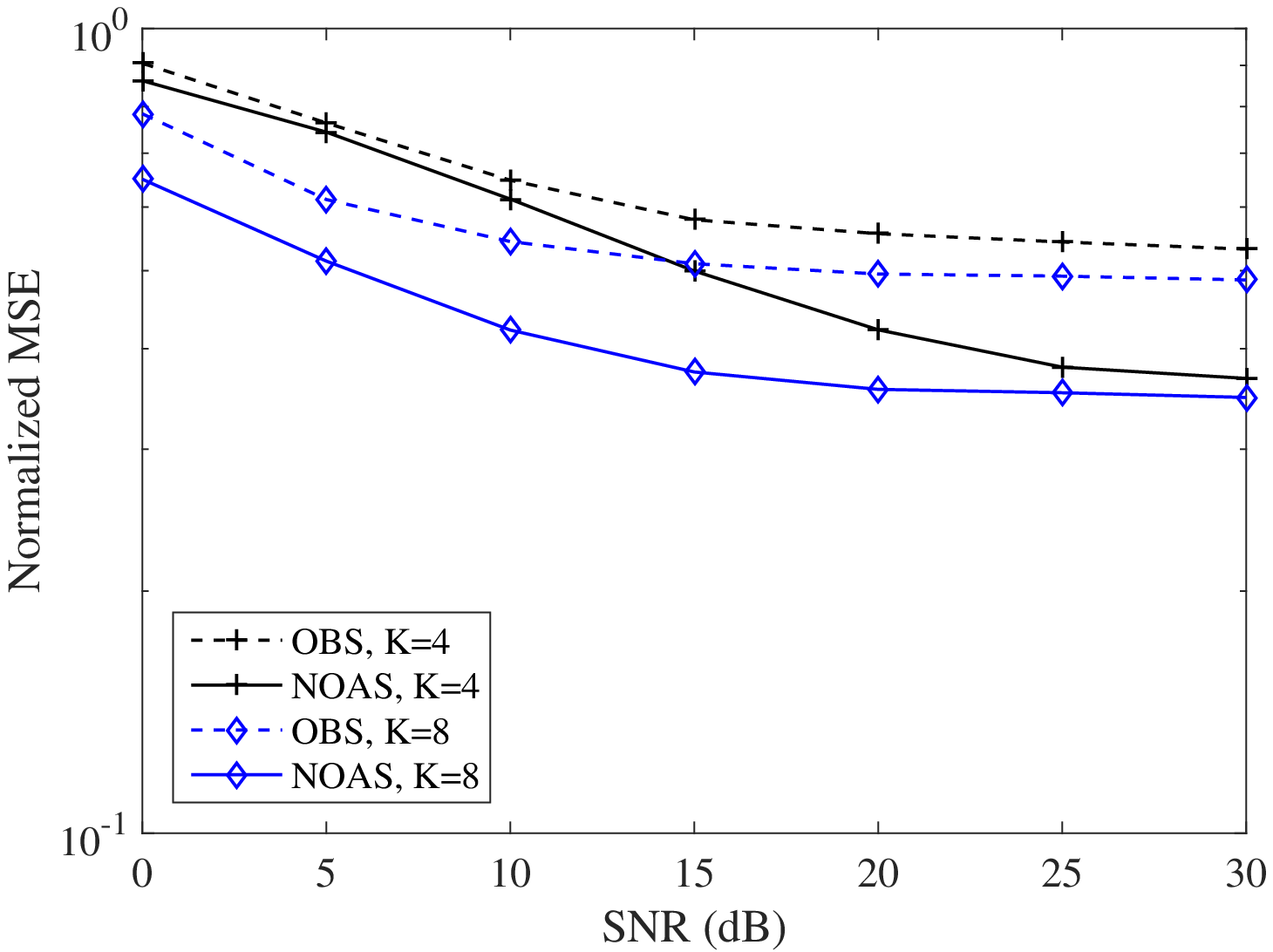}
\caption{Normalized MSE versus UL SNR, $\delta \theta=3^{\circ}$.}
\label{nmseA3}
\end{figure}

\begin{figure}[t]
\centering
\includegraphics[width=8.5cm]{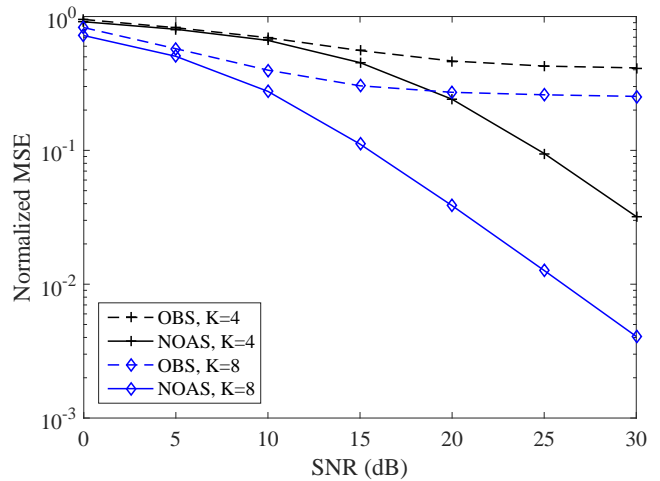}
\caption{Normalized MSE versus UL SNR, mmWave single path.}
\label{msemmwave}
\end{figure}

Fig. \ref{nmseA1} and Fig. \ref{nmseA3} demonstrate the normalized MSE (NMSE) of the proposed channel estimation for different $K$ with angle spread $\delta \theta=1^{\circ}$ and $\delta \theta=3^{\circ}$. It is seen that with the increase of SNR, the performances 
of both methods improve but will meet an error floor at a high SNR when the angular spread $\delta \theta=3^{\circ}$.
In all cases, the NOAS-based channel estimation can achieve a better NMSE than the OBS-based estimation because it forms more ``focusing'' beams towards the strongest directions of the users by sacrificing the orthogonality among users. Meanwhile, increasing $K$ also improves the NMSE performance.
The reason is that we assume $K$ spatial sampling in the first around and increasing $K$ will improve the accuracy of IMSB estimation, which in turn results in a better estimation performance.

Lastly, we consider the mmWave system, where the angle spread of channel reduces to almost zero, and the corresponding channel estimation performance is presented in Fig.~\ref{msemmwave}. It is seen from Fig. \ref{msemmwave} that the NOAS-based channel estimation tremendously outperforms the OBS-based method (or the conventional beamspace method). Compared to Fig. \ref{nmseA1} and Fig. \ref{nmseA3}, the NOAS-based channel estimation is much better under mmWave case since more channel power is focused on one single AOA, and accurately estimating this single AOA certainly yields better performance. For the same reason, the OBS-based one under mmWave case suffers from the severe power leakage, since it does not point exactly towards the user direction.

\section{Conclusions}
In this paper, we have proposed a novel hybrid transmission scheme for a MU massive MIMO system from the array signal processing perspective. Efficient channel estimation and AOA estimation algorithms under limited RF chains have been investigated. With the AOA information, the channel matrix can be  decomposed into an angle domain basis matrix and the corresponding angle domain channel matrix. We then present two ADMA schemes to serve multiple users with hybrid precoding, i.e., the orthogonal OBS and the non-orthogonal NOAS, where the former coincides with the exiting beamspace methods. 
It was shown that by pointing to the
strongest direction of the user, the NOAS scheme can alleviate power leakage effect and outperforms the beamspace methods.



\appendices
\section{Proof of Lemma 1}
From (\ref{eq:hkmatrix}), the orthogonality between two user channel vectors $\mathbf h_k$ and $\mathbf h_n$ can be measured by the correlation
\begin{eqnarray}
\gamma(k,n)& = & \frac{\mathbf h_k^{\mathcal H} \mathbf h_n}
{\norm {\mathbf h_k}\norm {\mathbf h_n}} \nonumber \\
& = & \frac{ \boldsymbol \alpha_k^{\mathcal H} \mathbf A_k^{\mathcal H}  \mathbf A_n \boldsymbol \alpha_n}
{\sqrt{\boldsymbol \alpha_k^{\mathcal H} \mathbf A_k^{\mathcal H} \mathbf A_k \boldsymbol \alpha_k}\displaystyle{\sqrt{\boldsymbol \alpha_n^{\mathcal H} \mathbf A_n^{\mathcal H} \mathbf A_n \boldsymbol \alpha_n}}} \nonumber \\
& = & \frac{\boldsymbol \alpha_k^{\mathcal H} \mathbf C_{k,n}\boldsymbol \alpha_n}
{\sqrt{\boldsymbol \alpha_k^{\mathcal H} \mathbf C_{k,k} \boldsymbol \alpha_k}\displaystyle{\sqrt{\boldsymbol \alpha_n^{\mathcal H} \mathbf C_{n,n} \boldsymbol \alpha_n}}},
\end{eqnarray}
where $\mathbf C_{k,n}=\mathbf A^{\mathcal H}_k \mathbf A_n$.

The $(i,j)$th entry of $\mathbf C_{k,n}$ can be written as
\begin{eqnarray}
C_{k,n}(i,j) &  = & {\mathbf a}^{\mathcal H} (\theta_{k,i}) {\mathbf a} (\theta_{n,j}) = \sum_{m=0}^{M-1} e^{jm2\pi \phi_{k,n}(i,j)} \nonumber \\ 
& = & \frac{1-e^{j2\pi M \phi_{k,n}(i,j)}}{1-e^{j2\pi \phi_{k,n}(i,j)}}, \label{ckn}
\end{eqnarray}
where
$\phi_{k,n}(i,j)= \frac{D}{\lambda} (\cos(\theta_{k,i})-\cos(\theta_{n,j}))$.

Since $\theta_{k,i}\ne \theta_{n,j}$ for $k\ne n$, $\phi_{k,n}(i,j)$ is not equal to $0$ for $k\ne n$. Therefore, $C_{k,n}(i,j)$ is bounded as $M\rightarrow \infty$, and $\displaystyle{\lim_{M\rightarrow \infty}\frac{C_{k,n}(i,j)}{M}}=0$.
Similar to (\ref{ckn}), we have $\displaystyle{\lim_{M\rightarrow \infty}\frac{C_{k,k}(i,j)}{M}}=0$ and $\displaystyle{\lim_{M\rightarrow \infty}\frac{C_{n,n}(i,j)}{M}}=0$ for $i\ne j$. Direct calculation yields that $C_{k,k}(i,i)=M$ and hence $\displaystyle{\lim_{M\rightarrow \infty}\frac{C_{k,k}(i,i)}{M}}=1$.

From the above discussion, we have
\begin{eqnarray}
\boldsymbol \alpha_k^{\mathcal H}\mathbf C_{k,k}\boldsymbol \alpha_k=M\sum_{i=1}^{P}|\alpha_{k,i}|^2+\sum_{i\ne j} \alpha^*_{k,i}\alpha_{k,j}C_{k,k}(i,j),
\end{eqnarray}
and
\begin{eqnarray}
\lim_{M\rightarrow \infty} \frac{\boldsymbol \alpha_k^{\mathcal H}\mathbf C_{k,k}\boldsymbol \alpha_k}{M}&=&\sum_{i=1}^{P}|\alpha_{k,i}|^2 \nonumber \\ 
&&+\lim_{M\rightarrow \infty}\sum_{i\ne j} \alpha^*_{k,i}\alpha_{k,j}\frac{C_{k,k}(i,j)}{M} \nonumber \\
&=&\sum_{i=1}^{P}|\alpha_{k,i}|^2.
\end{eqnarray}
Also, we have
\begin{eqnarray}
\lim_{M\rightarrow \infty}  \frac{\boldsymbol \alpha_k^{\mathcal H}\mathbf C_{k,n}\boldsymbol \alpha_n}{M}=\lim_{M\rightarrow \infty} \sum_{i,j} \alpha^*_{k,i}\alpha_{n,j}\frac{C_{k,n}(i,j)}{M}=0.
\end{eqnarray}
Therefore,
\begin{eqnarray}
\lim_{M\rightarrow \infty}\gamma(k,n) & = &
\lim_{M\rightarrow \infty}\frac{\boldsymbol \alpha_k^{\mathcal H} \mathbf C_{k,n}\boldsymbol \alpha_n}
{\sqrt{\boldsymbol \alpha_k^{\mathcal H} \mathbf C_{k,k} \boldsymbol \alpha_k}\displaystyle{\sqrt{\boldsymbol \alpha_n^{\mathcal H} \mathbf C_{n,n} \boldsymbol \alpha_n}}}  \nonumber \\
& = &
\lim_{M\rightarrow \infty}\frac{\displaystyle{\frac{\boldsymbol \alpha_k^{\mathcal H} \mathbf C_{k,n}\boldsymbol \alpha_n}{M}}}
{\displaystyle{\sqrt{\frac{\boldsymbol \alpha_k^{\mathcal H} \mathbf C_{k,k} \boldsymbol \alpha_k}{M}}}
\displaystyle{\sqrt{\frac{\boldsymbol \alpha_n^{\mathcal H} \mathbf C_{n,n} \boldsymbol \alpha_n}{M}}}} \nonumber \\
& = &
\frac{0}{\displaystyle{\sqrt{\sum_{i=1}^{P}|\alpha_{k,i}|^2}\sqrt{\sum_{j=1}^{P}|\alpha_{n,j}|^2}}}=0
\end{eqnarray}

\section{Derivation of (\ref{eq:gao2}) }
Since $\mathbf F \mathbf F^{\mathcal H} =\mathbf I$, we can rewrite $\mathbf A_k$ as
\begin{eqnarray*}
\mathbf A_k & = &  \mathbf F(\mathbf F^{\mathcal H}\mathbf A_k ) \nonumber \\
& = & \mathbf F[\mathbf F^{\mathcal H}\mathbf a(\theta_{k,1}),\ldots,\mathbf F^{\mathcal H}\mathbf a(\theta_{k,P})] \nonumber \\
& = & \mathbf F[\mathbf a_{k,1},\ldots,\mathbf a_{k,P}] \nonumber \\
&\approx &
\frac{1}{\sqrt{M}}
\left[
\begin{array}{ccccc}
& 1 & \ldots & 1 &  \\
\ldots &  e^{-j2\pi k_1\Delta f} & \ldots & e^{-j2\pi k_2\Delta f} & \ldots \\
\ldots & \vdots & \vdots & \vdots & \ldots\\
\ldots & e^{-j2\pi (M-1)k_1\Delta f} & \ldots & e^{-j2\pi (M-1)k_2\Delta f} & \ldots
\end{array}
\right] \nonumber \\
&&
\left[
\begin{array}{ccc}
\vdots & \vdots & \vdots \\
0 & \ldots & 0 \\
a_{k,1}(k_1) & \ldots & a_{k,P}(k_1) \\
\vdots & \vdots & \vdots \\
a_{k,1}(k_2) & \ldots & a_{k,P}(k_2) \\
0 & \ldots & 0 \\
\vdots & \vdots & \vdots
\end{array}
\right] \nonumber \\
& = & \tilde{\mathbf A}_k
\left[
\begin{array}{ccc}
\frac{a_{k,1}(k_1)}{\sqrt{M}} & \ldots & \frac{a_{k,P}(k_1)}{\sqrt{M}} \\
\vdots & \vdots & \vdots \\
\frac{a_{k,1}(k_2)}{\sqrt{M}} & \ldots & \frac{a_{k,P}(k_2)}{\sqrt{M}}
\end{array}
\right].
\end{eqnarray*}
From (\ref{eq:hkmatrix}), we have
\begin{eqnarray*}
{\mathbf h}_k  & = & \mathbf A_k \boldsymbol \alpha_k  \nonumber \\
& \approx & {\tilde {\mathbf A}}_k
\left[
\begin{array}{ccc}
\frac{a_{k,1}(k_1)}{\sqrt{M}} & \ldots & \frac{a_{k,P}(k_1)}{\sqrt{M}} \\
\vdots & \vdots & \vdots \\
\frac{a_{k,1}(k_2)}{\sqrt{M}} & \ldots & \frac{a_{k,P}(k_2)}{\sqrt{M}}
\end{array}
\right]
\left[
\begin{array}{c}
\frac{\alpha_{k,1}}{\sqrt{P}} \\
\vdots \\
\frac{\alpha_{k,P}}{\sqrt{P}}
\end{array}
\right] \nonumber \\
& = & {\tilde {\mathbf A}}_k {\tilde {\boldsymbol \alpha}}_k.
\end{eqnarray*}





\ifCLASSOPTIONcaptionsoff
  \newpage
\fi

\end{document}